\documentclass[letterpaper]{article}

\usepackage[T1]{fontenc}

\usepackage{geometry}
\usepackage{setspace}
\usepackage{xcolor}
\usepackage{url}

\usepackage[pagewise]{lineno}

\usepackage{achemso}
\setkeys{acs}{articletitle=true}

\usepackage{graphicx}
\usepackage{float}
\newfloat{scheme}{htbp}{los}
\floatname{scheme}{Scheme}
\floatname{chart}{Chart}
\newfloat{graph}{htbp}{loh}

\usepackage{chemformula} 
\usepackage[version = 4]{mhchem} 

\usepackage{authblk}

\author[1,*]{M. Saad~Bin-Alam}
\author[2,3]{Yunus~Denizhan Sirmaci}
\author[4]{Alejandro~Fernández-Hinestrosa}
\author[1]{Jianhao~Zhang}
\author[5,6]{Ksenia~Dolgaleva}
\author[5,6,7]{Robert~W. Boyd}
\author[4]{José Manuel~Luque-González}
\author[3]{Thomas~Pertsch}
\author[2,3]{Isabelle~Staude}
\author[1]{Jens~H. Schmid}
\author[1]{Pavel~Cheben}

\affil[1]{Quantum and Nanotechnologies Research Centre, National Research Council Canada, M-50, 1200 Montreal Road, Ottawa, ON, K1A 0R6, Canada}
\affil[2]{Institute of Solid State Physics, Friedrich-Schiller-University Jena, 07743 Jena, Germany}
\affil[3]{Institute of Applied Physics, Abbe Center of Photonics, Friedrich-Schiller-University Jena, 07745 Jena, Germany}
\affil[4]{Photonics and RF Research Lab, Telecommunication Research Institute (TELMA), University of Málaga, Bulevar Louis Pasteur 35, Málaga, 29010 Spain}
\affil[5]{School of Electrical Engineering and Computer Science, University of Ottawa, Ottawa, ON, K1N 6N5, Canada}
\affil[6]{Department of Physics, University of Ottawa, Ottawa, ON, K1N 6N5, Canada}
\affil[7]{The Institute of Optics and Department of Physics and Astronomy, University of Rochester, Rochester, New York 14627, USA}

\title{Directional and contra-directional coupling in Huygens' metawaveguide microring resonators}

\date{*Email: md.saad-bin-alam@nrc-cnrc.gc.ca}

\begin{document}

\maketitle

\begin{abstract}
\vspace{0.5em}
\hspace{-1.25cm}\begin{minipage}{0.4\textwidth}
{\includegraphics[width=2.5in]{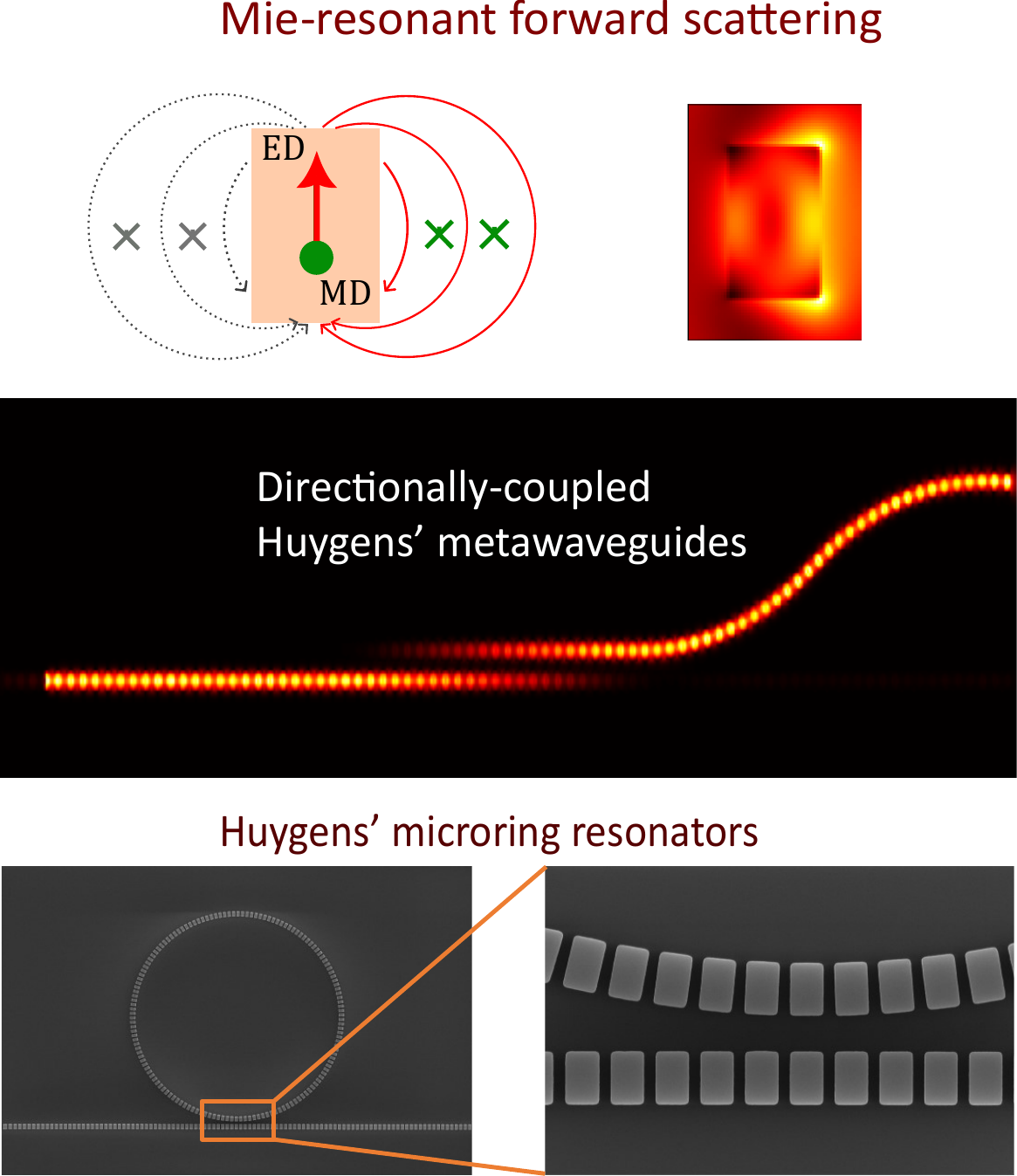}}
\end{minipage}
\begin{minipage}{0.6\textwidth}
Huygens' metawaveguides represent a transformative concept in photonic device engineering, enabling unprecedented control over light propagation. This study presents, for the first time, integrated Huygens'-based microring resonators and directional and contra-directional couplers, specifically designed for operation at the S- and C-band telecommunication wavelengths. By leveraging the unique properties of resonant Huygens' waveguides, we demonstrate efficient evanescent directional coupling with high-$Q$ resonators, characterized by negative group index and near-zero dispersion, which are critical for enhancing performance in compact, high-performance add-drop filters. The research further explores the implications of these novel structures on group index and group velocity dispersion, providing insights into their potential applications in nonlinear optics and quantum information technologies. Notably, the introduction of a hybrid subwavelength grating-Huygens' contra-directional coupler facilitates backward coupling between resonant and non-resonant metawaveguides, achieving a broad spectral rejection bandwidth. Our findings advance the integration of resonant metamaterials into scalable photonic platforms, laying the groundwork for innovative applications in optical communications, quantum photonics and sensing systems.
\section*{Keywords}
Huygens' scattering, dielectric nanoparticles, metawaveguides, integrated photonics, directional couplers, ring resonators
\end{minipage}
\end{abstract}



\section{Introduction}\label{section label}
Photonic metamaterials have fundamentally transformed our capacity to manipulate both free-space and guided optical waves, enabling functionalities far beyond those achievable with conventional dielectric media. Through precise engineering of their optical parameters, metamaterials provide unprecedented control over key properties of light—such as amplitude, phase, polarization, wavelength, and group velocity~\cite{zheludev2012metamaterials,smith2004metamaterials,shalaev2007optical,jahani2016all,yao2014plasmonic,monticone2017metamaterial,brener2019dielectric}. This control is typically realized by subwavelength-scale periodic modulation of materials with contrasting refractive indices~\cite{brener2019dielectric,soukoulis2011past,liu2008three,kadic20193d,chen2020flat,cheben2023recent,cheben2018subwavelength}. Moreover, by harnessing electromagnetic resonances within these periodic nanostructures, the light–matter interaction can be significantly amplified~\cite{kuznetsov2016optically,koshelev2020dielectric,staude2017metamaterial,bin2021ultra}.

Resonant metamaterials exhibit a range of distinctive functionalities, including tailored light transmission~\cite{decker2015high}, enhanced absorption~\cite{landy2008perfect}, control over propagation direction~\cite{park2021all,yu2011light}, beam shaping, focusing and lensing~\cite{kundtz2010extreme,khorasaninejad2016metalenses}, sensing~\cite{chen2012metamaterials}, and enhancement of nonlinear optical effects~\cite{lapine2014colloquium}. These capabilities are vital for diverse applications spanning classical and quantum information processing, communication, and signal routing~\cite{cortes2012quantum,stav2018quantum,solntsev2021metasurfaces}. 

Manipulation of the effective refractive index and group index is essential for optimizing mode coupling~\cite{son2018high,dulkeith2006group} and managing pulse propagation and dispersion in integrated photonics~\cite{gauthier2005slow,boyd2009controlling,turner2006tailored}. Such control enables advanced signal processing techniques indispensable to high-performance optical networks. Dispersion-engineered metamaterials, especially those incorporating subwavelength grating (SWG) structures that modulate the effective index~\cite{halir2015waveguide,halir2018subwavelength,bock2010subwavelength}, have emerged as key components in fiber-optic communication systems. These structures facilitate efficient fiber-to-chip coupling~\cite{cheben2010refractive,papes2016fiber},mode conversion~\cite{cheben2006subwavelength}, wavelength multiplexing~\cite{cheben2010refractive}, and high-performance directional and contra-directional couplers~\cite{halir2012colorless,lu2015broadband,wang2010uniform,shi2013silicon,shi2013ultra,naghdi2016silicon}, as well as micro-resonators~\cite{donzella2014sub,donzella2015design,eid2016fsr,naraine2023subwavelength} and other advanced photonic devices~\cite{cheben2023recent,cheben2018subwavelength}. 

In parallel, recent studies have unveiled remarkable optical phenomena — such as backward pulse propagation~\cite{liu2002superluminal,longhi2003superluminal,yanik2004time,stefaniuk2023nonlocality,he2022fast,zhang2021extraordinary}, pulse compression~\cite{peccianti2010subpicosecond,tan2010monolithic,blanco2014observation,tan2015nonlinear}, ultrafast signal processing and switching~\cite{chai2017ultrafast,wang2018chip,raja2021ultrafast}, and soliton and frequency-comb generation~\cite{del2007optical,brasch2016photonic,chang2022integrated} — enabled by waveguides exhibiting negative group index and anomalous dispersion~\cite{dogariu2001transparent,godey2014stability,kim2017dispersion}. Negative group index, wherein the phase velocity is opposite to the energy flow, gives rise to counterintuitive phenomena such as reversed pulse propagation and enhanced control of optical information. These regimes are accessible in resonant metamaterials~\cite{shcherbakov2015ultrafast,savelev2016solitary}.

While early implementations of resonant metawaveguides based on periodic Mie-scattering nanoparticles were hindered by excessive scattering and absorption losses~\cite{brener2019dielectric,savelev2016solitary,kuznetsov2016optically,bakker2017resonant}, recent progress has led to the realization of low-loss Huygens’ metawaveguides~\cite{sirmaci2023all,sirmaci2025ultracompact}. These structures exploit Huygens’ particles, which achieve backscattering suppression through constructive interference between co-aligned electric dipole (ED) and magnetic dipole (MD) resonances—satisfying the Kerker condition~\cite{kruk2017functional,staude2013tailoring,decker2015high,terekhov2017multipolar,kerker1983electromagnetic}. This condition ensures forward-directed radiation, enabling light propagation via near-field coupling between adjacent Huygens’ particles, while preventing scattering into radiation modes.

In this work, we present Huygens’ metawaveguide directional couplers and ring resonators designed to operate at the S- and C-band telecommunication wavelengths. These structures are used to characterize the group index and group velocity dispersion of resonant metawaveguides and to assess the performance of Huygens’ metawaveguide ring resonators as compact, high-performance add-drop filters. We further demonstrate efficient evanescent directional coupling between Huygens’ waveguides and high-$Q$ resonators engineered from metawaveguides with negative group index, near-zero dispersion, and anomalous dispersion profiles. Finally, we introduce a novel SWG–Huygens’ contra-directional coupler, enabling backward coupling between resonant and non-resonant metawaveguides while achieving a broad spectral rejection bandwidth.

\section{Results and Discussions}

\subsection{Huygens' nanoantennas and metawaveguides}

\begin{figure*}
\centering
{\includegraphics[width=1.0\linewidth]{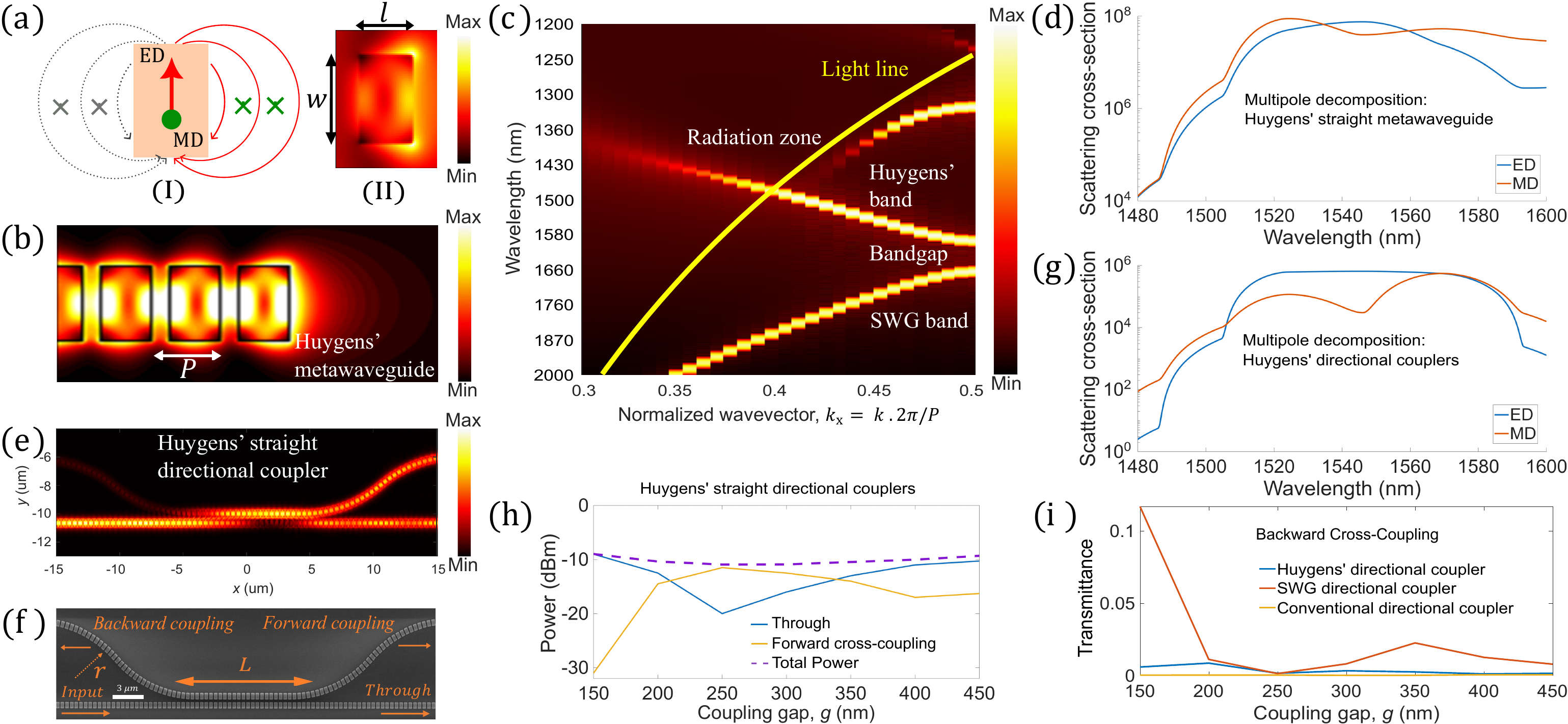}}
  \caption{(a) Overlap between Mie resonant electric and magnetic dipoles (ED and MD) in an individual dielectric nanoantenna (I) enables forward Huygens' scattering (II). (b) An array of Huygens’ antennas facilitates forward light propagation. (c) Dispersion band diagram of a Huygens’ metawaveguide shows the Huygens’ band around 1550 nm wavelength just below the light line (yellow line). (d) Individual antennas’ electric and magnetic dipole (ED and MD) scattering cross-section (in nm$^{2}$) in the straight metawaveguide for different normalized wavevectors, $k_{\mathrm{x}}$ ranging from 0.3 to 0.5 (2$\pi/P$) (the curves were fitted with the discrete data points extracted from the band diagram shown in (c). (e) Simulated light propagation and (f) a scanning electron micrograph (SEM) of a Huygens’ metawaveguide directional coupler with coupling gap, $g$ = 200 nm. (g) Individual antennas’ electric and magnetic dipole (ED and MD) scattering cross-section (in nm$^{2}$) in the forward cross-coupled waveguide (with coupling gap of 200 nm) for different normalized wavevectors, $k_{\mathrm{x}}$ ranging from 0.3 to 0.5 (2$\pi/P$). (h) Measured self and cross-coupling efficiency as the function of coupling gap extracted around 1530 nm wavelength (presented in Figure S3 (Part-A) in Supporting Information Section III). (i) Comparison between the cross-coupled backward propagation in Huygens’, SWG and conventional directional couplers as the function of coupling gap.}
  \label{fig:Figure1}
\end{figure*}

The core components of Huygens’ metawaveguides are individual dielectric nanoantennas that simultaneously exhibit Mie resonant electric and magnetic dipole (ED and MD) scattering within a desired spectral range~\cite{sirmaci2023all}. In Figure~\ref{fig:Figure1}a(I) the ED and MD resonances in a Huygens' particle are shown schematically, with the respective moments $p$ and $m$ satisfying the forward Kerker condition, \( p - \epsilon_{\mathrm{r}}m/c = 0 \) (where $\epsilon_{\mathrm{r}}$ is the medium’s relative permittivity and $c$ is the speed of light in vacuum), ensuring their comparable magnitudes and in-phase spectral
overlap~\cite{liu2018generalized,shamkhi2019transverse}. This in-phase excitation creates constructive interference of the ED and MD scattering in the forward direction and destructive interference in the backward direction, hence in zero backward scattering. Under this condition, the nanoantennas function as forward directional Huygens’ scatterers.

At the beginning to our design procedure, we carried out multipole decomposition analysis of an individual single-crystal silicon nano-cuboid antenna by exciting it with a total-field scattered-field (TFSF) source using Ansys Lumerical FDTD, and performed the decomposition of the total scattering cross-section using the generalized Kerker condition calculation~\cite{hinamoto2021menp}. In our simulation, we used a dispersive silicon model from the Lumerical material database, which was fitted based on silicon's refractive index, $n$ and the extinction coefficient, $k$ data shown in Figure S1(a-b) in Supporting Information Section I~\cite{palik1998handbook}. Through multipole decomposition analysis of the total scattering cross-section~\cite{alaee2015generalized} (see Figure S2(a) in Supporting Information Section II), we identified silicon nanoantenna with dimensions $l = 315$ nm, $w = 515$ nm, and $h = 220$ nm embedded in a silica cladding that acts as a unidirectional forward Huygens’ scatterer at the 1550 nm telecommunication wavelength. This is confirmed by the near-field mode profile shown in Figure~\ref{fig:Figure1}a(II). The height $h = 220$ nm was specifically chosen to ensure compatibility with the standard CMOS compatible silicon-on-insulator (SOI) waveguide platform. We carefully tuned the cuboid width-to-length ratio, $w:l$, for TE polarization to achieve the spectral overlap of the ED and MD resonances near the 1550 nm wavelength.

Next, we employed the optimized Huygens’ antennas to construct a Huygens’ metawaveguide, setting the antenna periodicity, $P$, and the antenna length, $l$, and introducing an inter-antenna gap, $g_{\mathrm{n}}$ (see Figure~\ref{fig:Figure1}(b)). The gap was optimized within the range 80-140 nm to maximize the combined forward-directional scattering efficiency of a dimer, composed of two consecutive nanoantennas. In this range, the fields of adjacent antennas remain sufficiently decoupled to preserve the fundamental scattering properties of individual antennas, while the proximity ensures the second antenna effectively captures the light scattered from the first antenna, enabling waveguiding effect. 

The calculated dispersion band of the infinitely long metawaveguide (see Figure~\ref{fig:Figure1}(c)) reveals two guided modes separated by the photonic band gap (PBG) around 1600 nm. The light line of SiO$_{\mathrm{2}}$ is indicated in yellow. The fundamental guided TE mode lies primarily in the long-wavelength region (below the PBG), referred to as the SWG band. Here, the chosen nanoantenna dimensions are considerably smaller than the wavelength of the propagating light. The SWG band extends up to the PBG at the edge of the Brillouin zone, where the array period equals half the effective wavelength. Below the PBG, the behavior of the SWG mode can be described by expressing the effective material refractive index, $n_{\mathrm{eff}}$, as a weighted average of the constituent refractive indices. The second guided mode, positioned above the PBG near the Huygens’ nanoantenna resonance wavelength, operates in the Huygens’ band and exhibits a negative group index~\cite{sirmaci2023all}. 

We chose the antenna periodicity, $P$ = 430 nm, to spectrally position the Huygens’ band near the 1550 nm telecommunication wavelength range and calculated the Bloch modes for a discrete set of normalized wavevectors, $k_{\mathrm{x}}$ ranging from 0.3 to 0.5 (2$\pi/P$). The chosen periodicity, $P$ of individual antennas with width, $w$ constitutes the duty cycle, $\delta$ = 0.73. Next, we calculated the field profiles of the Bloch modes ($|E|^{\mathrm{2}}$ from the middle of the $y$–$z$ cross-section of the antenna) of each branch at $k_{\mathrm{x}}$ = 0.414 around 1530 nm for the chosen periodicity, and launch the Bloch mode through the straight Huygens’ metawaveguide. We performed a multipolar decomposition analysis of the fields excited inside the individual antennas for the 1D metawaveguide periodic array. As shown in Figure~\ref{fig:Figure1}(d), multipole decomposition for the Huygens’ band revealed a good spectral overlap between ED and MD resonances, as required to fulfill the Kerker condition over a broadband (also provided in Figure S2(b) in Supporting Information Section II). The resonant properties of the coupled individual antennas are preserved in the array while the relative magnitudes of the electric and magnetic moments are locked across the Huygens' band. The antennas in the array effectively act as coupled Huygens’ scatterers over a wide spectral range of $\sim$60 nm~\cite{sirmaci2023all}. 

Following the same excitation technique described above, we performed the simulations of the Huygens’ directional couplers and the micro-resonators using the Ansys Lumerical FDTD software. In addition, we performed band diagram simulations of the SWG–Huygens’ contra-directional couplers using the MIT Photonic Bands (MPB) solver. Next, we fabricated the samples using a commercial foundry service, Applied Nanotools Inc. Fabrication involved electron-beam lithography on a commercial SOI substrate, comprising a 220 nm-thick single-crystalline silicon layer above a 3 $\mu$m buried oxide (BOX) layer on a silicon handle wafer. Electron-beam lithography was used to define waveguide patterns, which were subsequently transferred into the silicon layer through plasma etching. The samples were then coated with a 3 $\mu$m SiO$_{\mathrm{2}}$ cladding layer using plasma-enhanced chemical vapor deposition (PECVD). A second lithographic step and etching were performed to create the optical facets for light coupling, involving vertical etching through the top oxide, BOX layers, and into the silicon substrate. Lastly, the chips were diced near the etched facets. 

The fabricated samples were characterized using a continuous wave (CW) laser Agilent 8164B Lightwave Measurement System (LMS) with a linewidth of 0.01 nm operating in the wavelength range of 1460 nm to 1600 nm. To launch the input and collect and output light to measure the transmission, we used two spherical lensed fibers (radius of curvature: $R$ = 9 $\mu$m and mode field diameter: MFD = 3 $\mu$m); one was connected to the TE polarization-maintained laser source output, and another to the power meter. Every time we measured the transmission, we scanned the CW laser across its operating wavelength range to cover the transmission spectra of the ring resonators. The input and outputs of the bus waveguides were directly connected to the on-chip fiber-to-edge couplers~\cite{cheben2010refractive} to couple the source light and collect the transmitted signal. Coupling loss from optical fiber to waveguide was measured to be 1.6 dB/facet, and propagation losses of Huygens’ and conventional photonic wire waveguides were approximately 3 dB/cm and 1.5 dB/cm, respectively, determined from loss measurements of waveguides with varying lengths. Typical SWG waveguides have propagation losses around 2 dB/cm~\cite{cheben2010refractive}.

\subsection{Directional coupling of Huygens’ metawaveguides}

\begin{figure*}
\centering
{\includegraphics[width=1.0\linewidth]{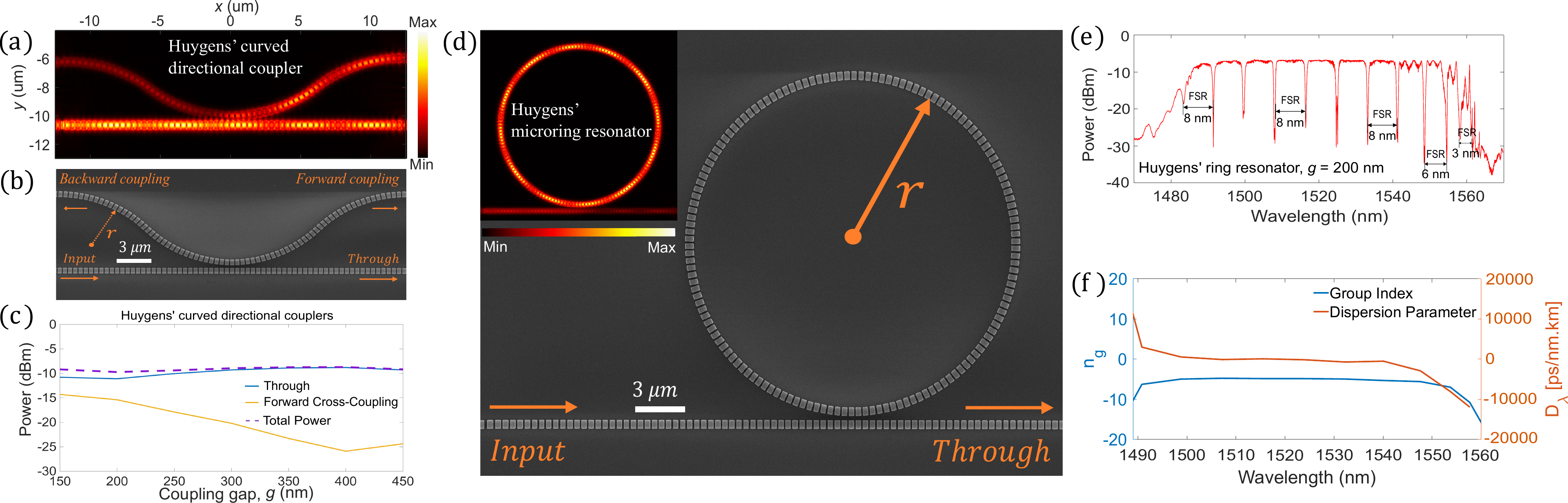}}
\caption{Simulated light propagation (a) and a scanning electron microscopy (SEM) image (b) of a straight-bend Huygens’ metawaveguide directional coupler with a coupling gap, $g$ = 150 nm and a bend radius, $r$ = 9 $\mu$m. (c) Measured self and cross-coupling efficiency as the function of coupling gap (see details in Figure S3 (Part-B) in Supporting Information Section III). (d) An SEM image of a Huygens ring resonator. The inset shows the simulated mode profile in the ring resonator. (e) The transmission spectra at through port of the Huygens' ring resonator with a coupling gap of 200 nm and a curved radius of 9 $\mu$m. (f) Group index, $n_{\mathrm{g}}$ and group velocity dispersion parameter, $D_{\mathrm{\lambda}}$ of the Huygens’ metawaveguides extracted from the measured resonance free-spectral range ($FSR$) of the ring resonator in (d) (presented in Figure S4 in Supporting Information Section IV).}
\label{fig:Figure2}
\end{figure*}

We designed and fabricated on-chip SOI directional couplers by placing two straight Huygens’ metawaveguides in parallel, separated by a coupling gap, $g$, to enable evanescent mode coupling, and finally measured the transmission power over the entire Huygens’ band from 1480 nm to 1560 nm wavelength for all the devices. The mode propagation in the directional coupler and the corresponding scanning electron microscopy (SEM) image of the device are shown in Figure~\ref{fig:Figure1}(e) and Figure~\ref{fig:Figure1}(f), respectively. The coupling length was designed as $L = 9\ \mu$m and the duty cycle to $\delta \approx 0.7$ to achieve 100$\%$ coupling efficiency~\cite{donzella2014sub}. 

Next, we performed the multipole decomposition analysis of the individual antennas located at the central position of the straight directional couplers' forward cross-coupled waveguide with varying coupling gap, $g$. Figure~\ref{fig:Figure1}(g) presents the simulated scattering cross-section for decomposed ED and MD moments excited by different wavelengths within the 1480-1560 nm Huygens' spectral range for different normalized wavevectors, $k_{\mathrm{x}}$ ranging from 0.3 to 0.5 (2$\pi/P$) for the coupling gap of 200 nm. The analysis for different $g$ from 150 - 300 nm are presented in Figure S2(c-h) in Supporting Information Section II. Compared to results for straight metawaveguides shown in Figure~\ref{fig:Figure1}(d), Figure~\ref{fig:Figure1}(g) suggests that in the directional coupler, MD amplitude varies across the central region of the Huygens' band, while ED amplitude remains approximately constant. 

Figure~\ref{fig:Figure1}(h) reveals that the coupling efficiency of the straight Huygens’ directional couplers strongly depends on the gap between the adjacent waveguides. Figure~\ref{fig:Figure1}(h), summarizing the data in Figure S3 (Part-A) in Supporting Information Section III, shows that at a 150 nm gap the 9 µm straight coupler exhibits minimal net power transfer because strong coupling causes light to cross-couple and subsequently couple back to the through waveguide. Increasing the gap to 200 nm sharply raises cross-coupled power, making the through and drop powers comparable—$i.e.$, a 3-dB directional coupler. At 250 nm, most power transfers to the cross-coupled port, approaching 100\% coupling. This rapid variation continues up to 400 nm and stabilizes near 450 nm (the weakly coupled regime).

Since Huygens' scatterers are designed to suppress back-scattering (see Figure~\ref{fig:Figure1}(a)), they effectively reduce propagation losses associated with back-reflections in periodic resonant metawaveguides. Our simulated results confirm that back-reflection remains suppressed even in coupled Huygens’ metawaveguides, where energy exchange between grating structures could otherwise induce back-scattering~\cite{li2016backscattering}. Comparisons with a non-resonant SWG metawaveguide directional coupler with the same coupling length ($L$ = 9 $\mu$m), duty cycle ($\delta$ = 0.72), antenna width ($w$ = 180 nm), and periodicity ($P$ = 250 nm) show that  backward directional coupling remains lower in the coupled Huygens’ waveguides compared to the SWG waveguides, as shown in Figure~\ref{fig:Figure1}(i).

\subsection{Huygens’ ring resonators}

\begin{figure*}
\centering
{\includegraphics[width=1\linewidth]{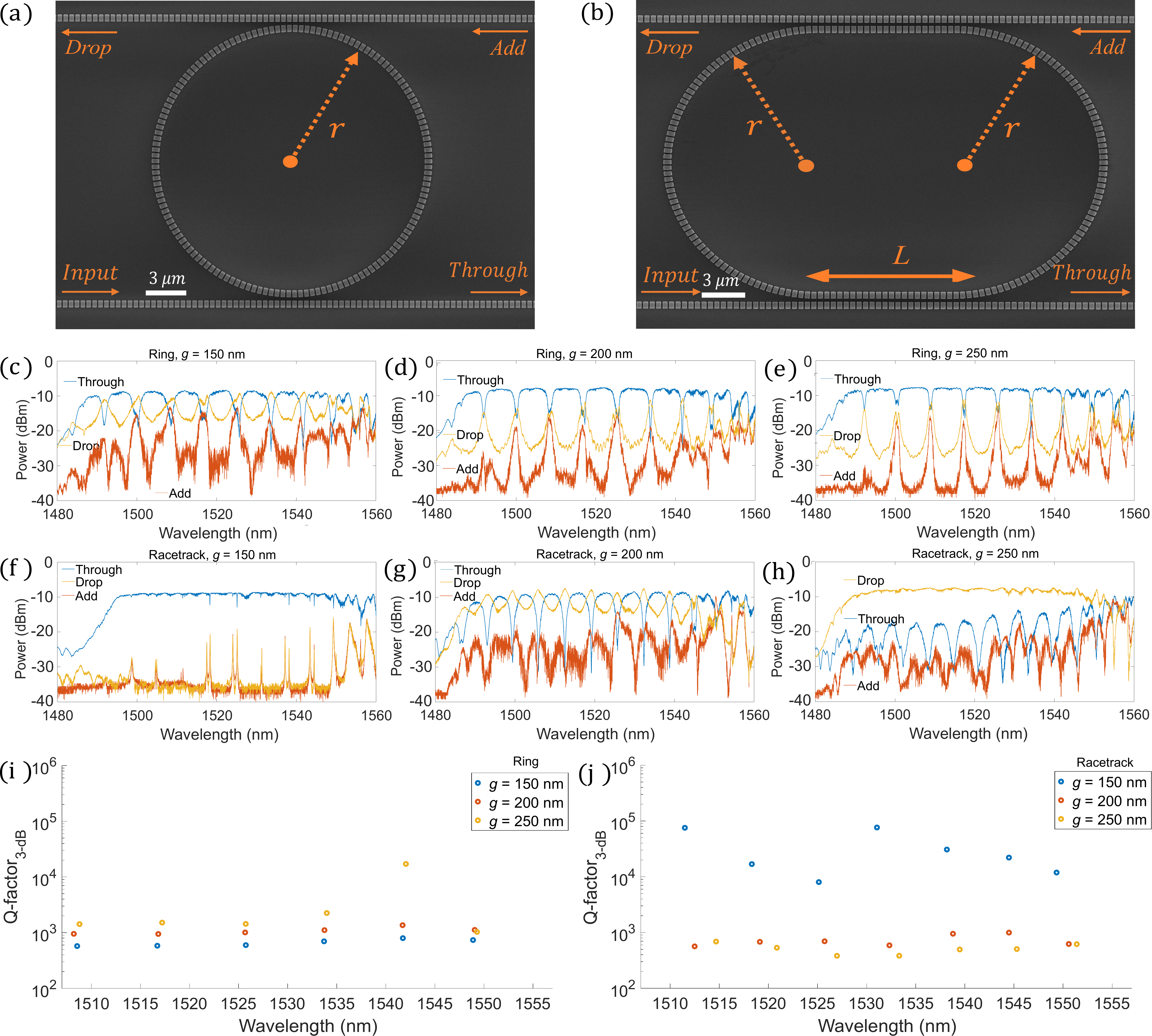}}
\caption{(a-b) Scanning electron micrographs (SEM) of the Huygens’ four-port ring and racetrack resonator add-drop filters. (c-e) Measured transmission spectra corresponding to the through, add, and drop ports of the ring resonator add-drop filter shown in (a) for  the coupling gap of 150 nm, 200 and 250 nm, respectively. (f-h) Measured transmission spectra corresponding to the through, add, and drop ports of the racetrack resonator add-drop filter shown in (b)
for the coupling gap of 150 nm, 200 and 250 nm, respectively. (i-j) Measured Q-factors at 3-dB bandwidth of the resonances corresponding to the ring and racetrack resonator add-drop filters shown in (c-e) and (f–h) for different coupling gaps, respectively.}
\label{fig:Figure3}
\end{figure*}

Next, we substituted one of the two straight Huygens’ metawaveguides of the directional coupler with a curved metawaveguide. To investigate the cross-coupling characteristics, we analyzed such straight-curved metawaveguide directional couplers by varying the coupling gap, while keeping the bend radius fixed at $r = 9\ \mu$m to minimize bend loss. Figure~\ref{fig:Figure2}(a) depicts the simulated mode propagation of the curved directional coupler with a coupling gap of 150 nm, and Figure~\ref{fig:Figure2}(b) shows a corresponding SEM image of a fabricated structure. Figure~\ref{fig:Figure2}(c), summarizing the measured data presented in Figure S3 (Part-B) in Supporting Information Section III, suggests that the cross-coupling power transfer as a function of the coupling gap varies significantly more slowly for the curved directional couplers compared to the straight directional couplers. These different characteristics arise from the effectively reduced coupler length for the curved couplers. The power conservation in the straight and curved directional couplers is discussed in Supporting Information Section VIII.


We utilized the straight-bend metawaveguide directional coupler to construct a Huygens’ ring resonator. Figure~\ref{fig:Figure2}(d) shows an SEM image of the fabricated ring resonator with the radius of 9 $\mu$m coupled to a bus metawaveguide via a gap of 200 nm. The inset shows the calculated resonator electric field at a 1550 nm wavelength. The transmission spectra near the Huygens’ band, presented in Figure~\ref{fig:Figure2}(e), reveal that the free-spectral range, $FSR$, of the excited resonance modes remains almost constant in the central spectral zone, but gradually decreases as the wavelength approaches the photonic bandgap (PBG). The measured $Q$-factor of the resonances across the Huygens’ band is around 2000. 

Using the FSR corresponding to the resonances shown in Figure~\ref{fig:Figure2}(e), we extracted the metawaveguide group index, $n_{\mathrm{g}}$, from the relation $FSR = \delta\lambda_{\mathrm{FSR}} = \frac{\lambda^{2}}{n_{\mathrm{g}} L_{\mathrm{o}}}$, where $\lambda = \frac{n_{\mathrm{eff}}L_{\mathrm{o}}}{m}$ is the resonance wavelength, $n_{\mathrm{eff}}$ is the metawaveguide effective index, $L_{\mathrm{o}} = 2\pi r$ is the ring circumference, and $m =$ 1, 2, 3,... is the integer of the mode number. Due to the inverse frequency-wavevector dispersion relation (shown in Figure~\ref{fig:Figure1}(c)), the group index, $n_{\mathrm{g}}$, for the Huygens' band is negative. The flat and descending $FSR$ depicted in Figure~\ref{fig:Figure2}(e) correspond to zero and anomalous group velocity dispersion (GVD), respectively. The dispersion parameter, $D_{\mathrm{\lambda}} = \frac{\partial}{\partial\lambda} \frac{1}{v_{\mathrm{g}}}$, where $v_{\mathrm{g}} = -c/n_{\mathrm{g}}$ and $c$ is the light velocity in free space, also follows this behavior. Figure~\ref{fig:Figure2}(f) depicts the group index, $n_{\mathrm{g}}$, averaging -4.65 in the central zone of the Huygens' band, decreasing to -10 near the PBG edges. In contrast, the dispersion parameter, $D_{\mathrm{\lambda}}$, in the same plot demonstrates near zero dispersion in the central zone which eventually declines to -10,000 ps/nm·km as the photonic bandgap is approached. Such feature is unique for the Huygens' metawaveguide ring resonators, as shown in the comparison with the conventional and the SWG ring resonators presented in Figure S4 in Supporting Information Section IV.

\subsection{Characteristics of Huygens' micro-resonator add-drop filters}

We also studied four-port add-drop filter resonators using the Huygens’ directional couplers demonstrated earlier. Figure~\ref{fig:Figure3}(a) and Figure~\ref{fig:Figure3}(b) present SEM images of the fabricated double bus-coupled ring and racetrack resonators, each featuring input, through, add, and drop ports. In both designs, the radii of the curved sections are set to $r$ = 9 $\mu$m, while the straight coupling length of the racetrack resonators is $L$ = 9 $\mu$m. Figure~\ref{fig:Figure3}(c-e) show the experimentally observed transmission spectra across the Huygens’ band at the through, drop, and add ports in the ring resonator filters for coupling gaps of 150 nm, 200 nm, and 250 nm. Similarly, Figure~\ref{fig:Figure3}(f-h) present the corresponding spectra for the racetrack resonators for the same coupling gaps.

In the double bus-ring resonator filters, the observed resonances exhibit over-coupled, critically coupled, and under-coupled characteristics as the coupling gap is initially set at 150 nm (see Figure~\ref{fig:Figure3}(c)) and then gradually increased up to 250 nm by 50-nm step (see Figure~\ref{fig:Figure3}(d-e)). The extinction ratio of the through-port resonances gradually decreases with increasing coupling gap, as the coupling strength is reduced following the coupling condition trend exhibited by the corresponding curved directional couplers with the same coupling gap levels in Figure~\ref{fig:Figure2}(c) and in Figure S3 (Part-B) in Supporting Information Section III. For both over- and critically coupled regimes shown in Figure~\ref{fig:Figure3}(c) and Figure~\ref{fig:Figure3}(d), the insertion loss difference between the add and drop ports of the ring-based filters is approximately 5 dB. In the under-coupled regime shown in Figure~\ref{fig:Figure3}(e) when the coupling gap is 250 nm, the resonances exhibit reduced peak-to-peak insertion loss difference between the power coming out from the add and drop ports. 


Unlike ring-resonator add–drop filters, double bus–racetrack resonator filters show distinct gap-dependent coupling trends (Figure~\ref{fig:Figure3}(f–h)). Comparison with Figure~\ref{fig:Figure1}(h) and Figure S3 (Part-A) in Supporting Information Section III shows they follow the same behavior as straight directional couplers with identical coupling lengths and gaps. Here, the bus–racetrack power coupling varies more rapidly than in bus–ring filters, yielding minimal net transfer at the initial 150 nm gap due to strong coupling (see Figure~\ref{fig:Figure3}(f)). Far from critical coupling, the through-port resonance extinction ratio is 2–5 dB, making add- and drop-port insertion losses comparable. At the increased gap of 200 nm, the extinction ratio increases to 15–20 dB (see Figure~\ref{fig:Figure3}(g)). A further 50 nm gap increment reduces the through-port transmission to 10 dB below the drop port (see Figure~\ref{fig:Figure3}(h)). This rapid cross-coupling oscillation is consistent with straight directional coupler behavior. The power conservation in the ring and racetrack resonators is discussed in Supporting Information Section VIII.

In the weakly coupled ring resonator filter, some resonances exhibit spectral splitting (see Figure~\ref{fig:Figure3}(e)), which becomes more pronounced at larger coupling gaps (see Figure S5 in Supporting Information Section V). For the racetrack resonator with a 150 nm gap (weak power coupling), many resonances show strong splitting with separation exceeding the 3-dB bandwidth (see Figure~\ref{fig:Figure3}(f)). This asymmetric splitting in both filters under reduced net power coupling indicates possible back-reflection effects~\cite{bogaerts2012silicon,li2016backscattering,zhang2008resonance}, attributed to light scattering from nanoantenna array. Although Huygens’ antennas eliminate back-scattering in ideal straight metawaveguides, in bent waveguides~\cite{sirmaci2023all} some scattering loss is unavoidable. Therefore, back-scattering cannot be neglected in ring resonators or in the curved sections of racetrack resonators. Back-scattered signal can build up in ring or racetrack resonators, enhancing counter-clockwise mode propagation and causing the observed resonance splitting, particularly in weakly coupled devices.




Finally, we examine the resonance $Q$-factors of the add–drop filters. Figure~\ref{fig:Figure3}(i) shows that the 3-dB $Q$-factors of ring-based filters increase with larger coupling gaps, entering the weakly coupled regime. Their average $Q$ remains around $10^{3}$, while sharper split resonances exhibit higher values. Similarly, the split resonances of racetrack-based filters with the smallest coupling gap—corresponding to weak power cross-coupling (see Figure~\ref{fig:Figure1}(h))—show high $Q$-factors of $10^{4}$–$10^{5}$, up to two orders of magnitude larger than those of racetrack filters with larger gaps (see Figure~\ref{fig:Figure3}(j)).

\subsection{Add-drop filtering with contra-directional couplers}

\begin{figure*}
\centering
{\includegraphics[width=1\linewidth]{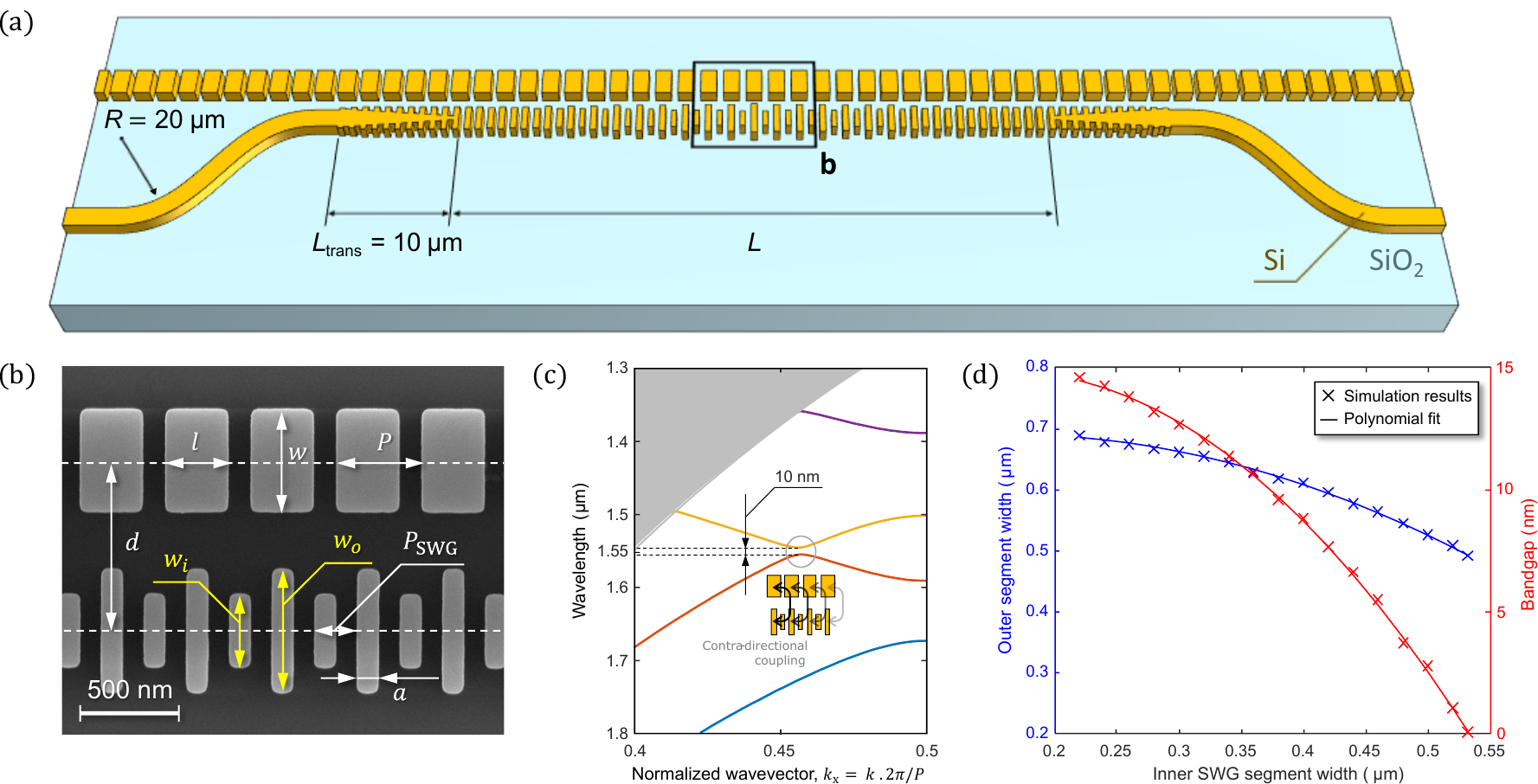}}
  \caption{(a) Schematic 3D view of the SWG-assisted Huygens’ contra-directional coupler (CDC). (b) SEM image of the SWG-assisted Huygens’ CDC. The period of the lateral SWG waveguide is exactly half the period of the Huygens’ waveguide. Modulating the width of the SWG waveguide segments enables the backward coupling of light from the Huygens’ waveguide. (c) Band diagram corresponding to the SEM image in (b). The SWG waveguide was tuned to enable energy transfer to the Huygens’ waveguide at the center of its operating bandwidth ($\lambda$ = 1550 nm). (d) Outer segment width ($w_\mathrm{o}$) as a function of inner segment width ($w_\mathrm{i}$) of the lateral SWG waveguide, which together modify the bandgap width while keeping the center wavelength constant at 1550 nm (blue data). Bandgap as a function of inner segment width ($w_\mathrm{i}$) of the SWG waveguide (red data). Markers represent simulation results from 3D band diagram calculations, while solid curves are second-degree polynomial fits to the simulation data.}
  \label{fig:Figure4}
\end{figure*}

Contra-directional couplers (CDC) are optical metamaterial structures that enable wavelength-specific power exchange between different waveguide supermodes traveling in opposite directions. They function as wavelength filters, with the backward-coupled fields directed to a port distinct from the input. When integrated into silicon photonics platforms, CDCs provide a compact and effective solution for on-chip spectral control of light. In this section of the article, we design and experimentally demonstrate a CDC by side-loading the Huygens’ waveguide with a modulated subwavelength grating (SWG) waveguide, as shown in Figure~\ref{fig:Figure4}(a). The use of a SWG waveguide enables considerable interaction between the supermodes of the structure while preserving the radiation pattern of the Huygens’ waveguide segments. We then experimentally assess optical add-drop filters comprising both a Huygens’ directional coupler and a Huygens’ CDC, demonstrating that the latter provides spectral control over the racetrack resonances.

To design the CDC, we first set the period of the SWG waveguide to $P_\mathrm{SWG}$ = 215 nm, which is precisely half the period of the Huygens’ waveguide ($P$ = 430 nm), as shown in Figure~\ref{fig:Figure4}(b). The duty cycle of the SWG waveguide was set to $\delta = a/P_\mathrm{SWG} \sim$ 0.5, ensuring equal dimensions for both the silicon blocks and gaps ($a$ $\sim$ 108 nm). This balance facilitates fabrication by avoiding overly small features. The optimal center-to-center separation between the Huygens’ waveguide and the SWG waveguide was determined to be $d$ = 850 nm, maximizing evanescent field coupling while conserving the radiation pattern of the Huygens’ waveguide segments. 

By initially setting the width of the SWG waveguide segments equal to the width of the Huygens’ waveguide segments, we observed a small bandgap of approximately $\Delta\lambda_\mathrm{init} \sim$ 1.5 nm at the center of the Huygens’ waveguide operating bandwidth ($\lambda$ = 1550 nm). This bandgap corresponds to the contra-directional coupling between the Huygens’ waveguide and the SWG waveguide supermodes and was analyzed using band diagram calculations, as shown in Figure~\ref{fig:Figure4}(c). The contra-directional coupling is a power exchange enabled by the grating vector, and occurs only for a specified spectral region--the “bandgap” (grey circle). A detail of the contra-directional coupling mechanism in the band diagram is provided in Figure S6 in Supporting Information Section VI. To increase the bandgap, we narrowed one segment in each pair of silicon segments in the SWG waveguide, $i.e.,$ we made the inner segment width ($w_\mathrm{i}$) smaller. This change induced a blue-shift in the bandgap center wavelength, attributed to the reduced silicon content thus a lower equivalent refractive index in the SWG waveguide. To shift the bandgap center back to the target wavelength ($\lambda$ = 1550 nm), we optimally widened the adjacent segments, $i.e.,$ we made the outer segment width ($w_\mathrm{o}$ $>$ $w_\mathrm{i}$) larger. After repeating this procedure for each inner segment width in the range $w_\mathrm{i}$ $\mathrm\epsilon$ [0.2,0.55] $\mu$m, we obtained the simulation results shown in Figure~\ref{fig:Figure4}(d). The maximum achievable bandgap was found to be of approximately $\Delta\lambda_\mathrm{max} \sim$ 15 nm. Both the bandgap as a function of inner segment width, $\Delta\lambda$($w_\mathrm{i}$), and the outer segment width as a function of inner segment width, $w_\mathrm{o}$($w_\mathrm{i}$), were fitted with second-degree polynomials, providing the basis for the design of an apodized filter.

\begin{figure*}
\centering
{\includegraphics[width=1\linewidth]{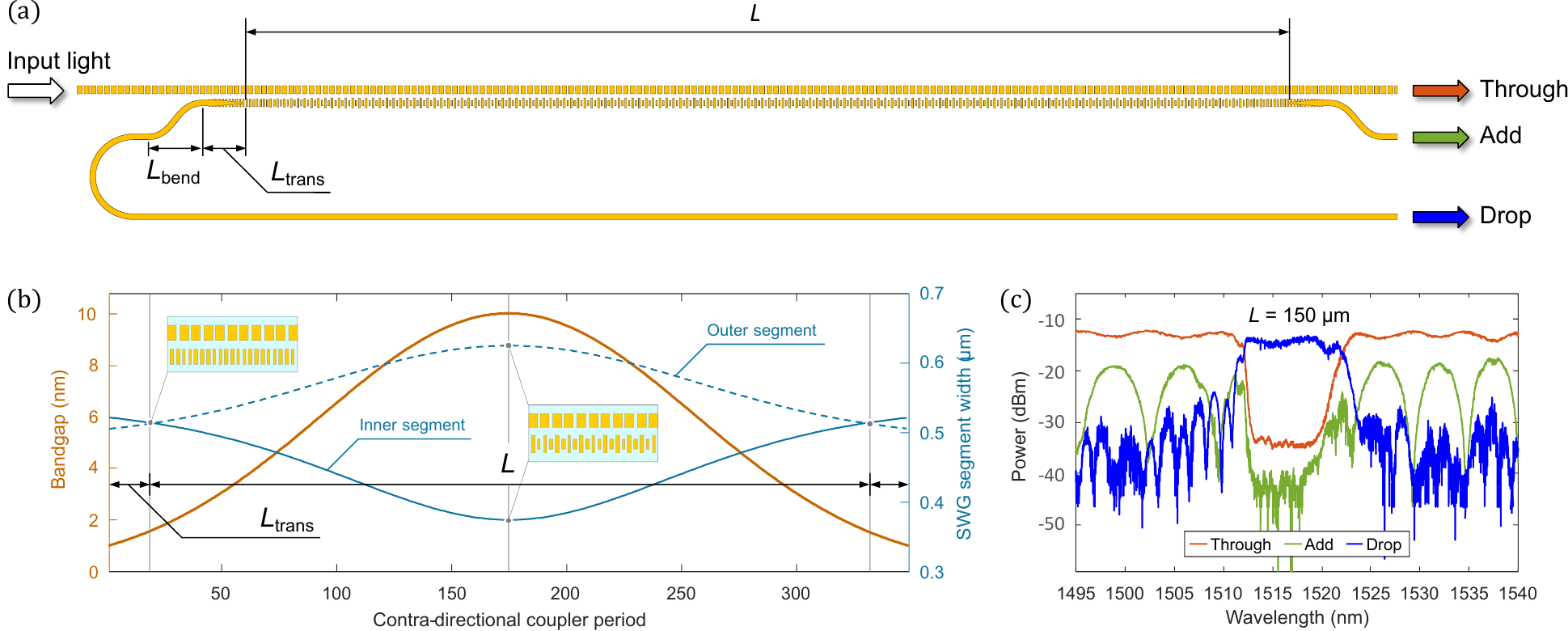}}
  \caption{(a) Schematic of the apodized SWG-assisted Huygens’ CDC. Light is injected at the input port (left) and optically measured at the through, add and drop ports (right). The total length of the device consists of the length of the CDC ($L$), the length of the SWG waveguide to solid waveguide transitions ($L_\mathrm{trans}$), and the length of the solid waveguide bends ($L_\mathrm{bend}$). (b) Target Gaussian apodization of the CDC bandgap (orange curve). The edges of the apodization function ($\sim$1 nm) are set to 10 $\%$ of the peak bandgap ($\Delta\lambda_\mathrm{peak}$ = 10 nm). Inner and outer segment widths (blue curves) are mapped through the inverse of the function $\Delta\lambda$($w_\mathrm{i}$) and the function $w_\mathrm{o}$($w_\mathrm{i}$), respectively. (c) Experimental optical measurements of the through, add and drop ports when light is injected from a tunable laser to the input port of the device. The target peak bandwidth closely matched the band diagram calculations, but the center wavelength experienced a blue-shift due to fabrication-induced feature deviations.}
  \label{fig:Figure5}
\end{figure*}

\begin{figure*}
\centering
{\includegraphics[width=1\linewidth]{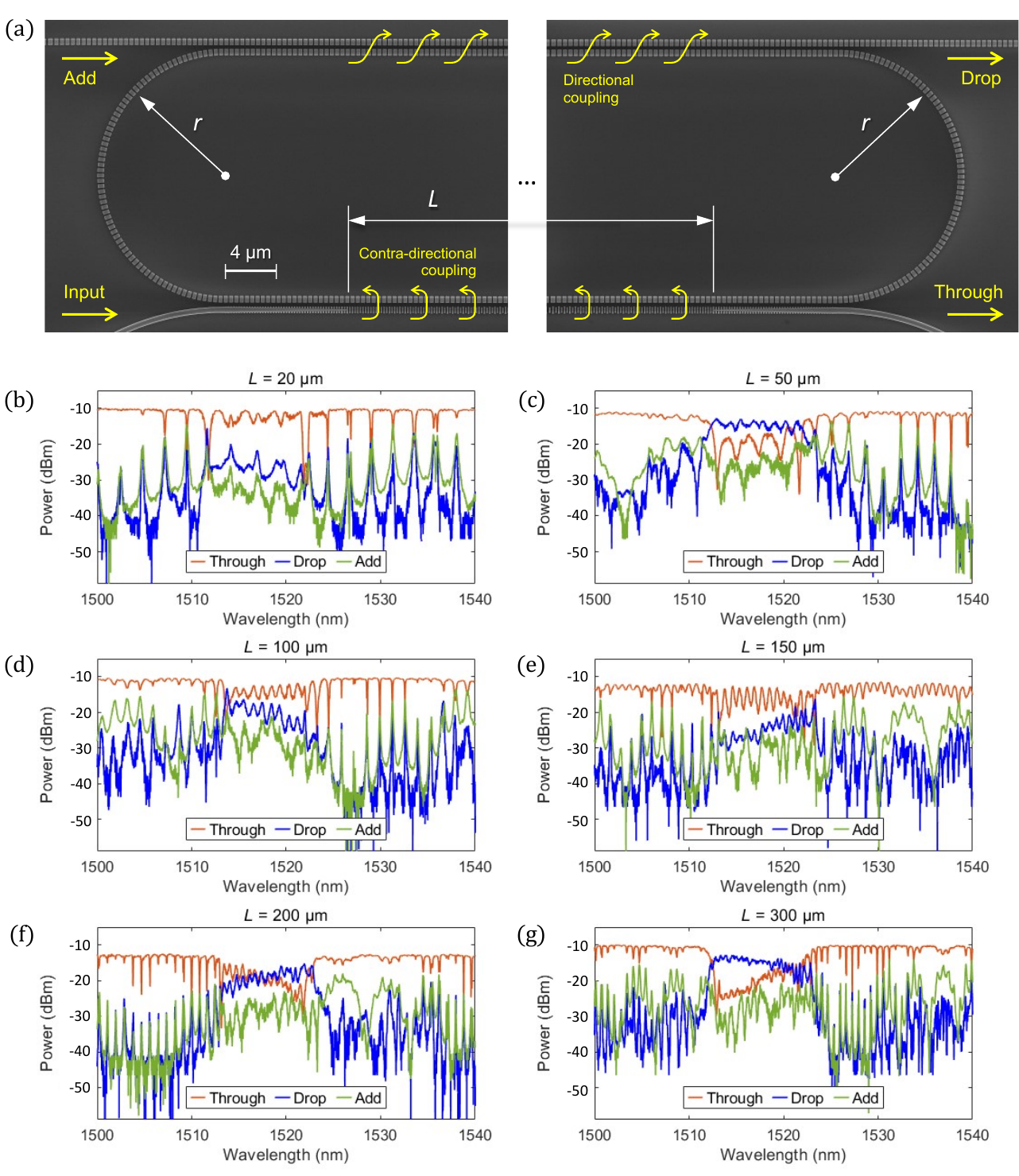}}
  \caption{(a) SEM image and port description of the CDC-loaded racetrack resonator. (b-g) Experimentally measured optical power at the through, add and drop ports of the CDC-loaded racetrack resonator for varying CDC lengths. The $FSR$ of the resonances are 2.3, 1.8, 1.3, 0.9, 0.6 and 0.5 nm, respectively.}
  \label{fig:Figure6}
\end{figure*}

Building on the previous bandgap calculations, we designed an apodized single-band filter with a peak bandwidth of $\Delta\lambda_\mathrm{peak}$ = 10 nm. The initial length of the CDC was set to $L$ = 150 $\mu$m, which for a bandgap of 10 nm provides a calculated rejection of 30 dB, and a Gaussian apodization window was applied, with the grating edges reflecting 10$\%$ of the peak bandwidth ($i.e.,$ 1 nm). Assuming a group index of ng = 4, the design targets a theoretical rejection ratio exceeding 60 dB to ensure efficient contra-directional power transfer to the drop port. The Gaussian apodization of the grating is employed to achieve low insertion loss and reduced out-of-band ripples. The filter comprises three distinct sections, as shown in Figure~\ref{fig:Figure5}(a). In the central section of the device, corresponding to the apodized CDC, the mode traveling to the right through the Huygens’ waveguide is coupled backward to the SWG waveguide. The backward coupling occurs within a 10-nm wide band centered at $\lambda$ = 1550 nm. The SWG waveguide segments are modulated based on the data shown in Figure~\ref{fig:Figure4}(d) and following the target Gaussian bandgap profile. The results of these calculations are shown in Figure~\ref{fig:Figure5}(b). After coupling to the SWG waveguide, the mode transitions into a 450-nm-wide strip waveguide over a length of $L_\mathrm{trans}$ = 10 $\mu$m, $i.e.$, 46 periods of the SWG waveguide. The size of the SWG inner and outer segments is continuously adjusted to close the bandgap as needed by the Gaussian apodization. At the junction with the SWG–wire coupler, the SWG waveguide has the same inner and outer segment widths. The strip waveguide then bends away from the Huygens’ waveguide over a length of $L_\mathrm{bend}$ = 35 $\mu$m and using a bending radius of R=20 $\mu$m. Identical structures are also included at the output of the SWG waveguide to extract any directionally coupled light. The experimental results of the filter are shown in Figure~\ref{fig:Figure5}(c). The bandwidth remained fixed at 10 nm after fabrication, and the power loss was around 1 dB when referenced to the straight Huygens’ waveguide transmittance. We attribute the observed blue-shift of the center wavelength to deviations in the layout feature sizes during fabrication. An interference pattern was observed at the add port, which results from residual directional coupling between the Huygens’ and the SWG waveguides. However, this interference had minimal impact on the transmittance at the through port.

To evaluate the influence of the CDC on racetrack-resonator performance, devices with different CDC lengths were fabricated and characterized experimentally. A device under test is shown in the SEM images of Figure~\ref{fig:Figure6}(a). The coupling length was varied from 20 $\mu$m to 300 $\mu$m in several steps, which in turn modified the $FSR$ of the resonances from 2.3 to 0.5 nm, respectively. A more detailed view of the racetrack resonances is provided in Figure S7 in Supporting Information Section VII. The transmission spectra of the CDC-loaded racetrack resonators at the through, add and drop ports are presented in Figure~\ref{fig:Figure6}(b-g). The observed resonances exhibited an exceptionally low $FSR$, while the CDC also limited the light coupling spectrum of the racetrack to a 10-nm rejection band centered within the Huygens’ operation bandwidth. This demonstrated the ability to tailor the Huygens’ racetrack resonances using an envelope function provided by an apodized SWG-assisted CDC. The power conservation in the CDC and the CDC-loaded racetrack resonators is discussed in Supporting Information Section VIII.

\section{Conclusion}

In summary, this work has introduced and experimentally validated a new class of integrated photonic components based on Huygens’-resonant microring resonators, and directional and contra-directional couplers, all engineered for operation at the S- and C-band telecommunication wavelengths. By exploiting the unique dispersion and group velocity properties of Huygens’ waveguides, we demonstrated efficient evanescent coupling in micro-ring resonators implemented in both directional and contra-directional configurations, relevant for compact, high-performance add-drop filters. We also explored the broader implications of these structures for controlling group index and group velocity dispersion, pointing to promising applications in nonlinear and quantum photonics. A key innovation is the development of a hybrid subwavelength grating–Huygens’ contra-directional coupler, enabling backward coupling across resonant and non-resonant metawaveguides and supporting broad spectral rejection bandwidths. This advancement facilitates the design of FSR-free racetrack resonators, opening new pathways for spectral engineering. 

Our Huygens’ metawaveguide devices differ from conventional photonic-crystal (PhC) waveguides in that light guidance and dispersion arise from locally resonant dielectric antennas rather than from extended Bloch modes supported by a large periodic lattice. In PhCs, phenomena such as slow light, negative group velocity, and contra-directional coupling depend on long-range Bragg interference across many unit cells, making them sensitive to fabrication disorder and difficult to tune after fabrication. In contrast, Huygens metawaveguides employ directionally scattering electric–magnetic resonators whose near-field coupling can be engineered locally. This opens a path towards a potentially reduced device footprint, flexible dispersion control, improved tolerance to fabrication imperfections, and dynamic reconfigurability, making Huygens architectures more versatile than traditional PhC-based waveguide structures.

The Huygens’ metawaveguide devices operate over a bandwidth approaching 100 nm, which is fundamentally limited by the spectral overlap of the ED and MD resonance linewidth. In future work, this bandwidth could be further extended by engineering the antenna geometry to simultaneously excite higher-order electric and magnetic quadrupole resonances (EQ and MQ) within a comparable wavelength range, thereby broadening the effective Huygens regime. Additional bandwidth-limiting phenomena arise from the photonic bandgap at low frequencies and the onset of diffraction losses at high frequencies near the light line, which can be optimized through judicious design. In addition to bandwidth enhancement, further improvements in add–drop filter performance could possibly be realized by increasing the radii of the resonators’ curved sections to mitigate bend-induced scattering losses and enhance overall extinction ratio and transmission efficiency. 

Altogether, we anticipate that this work will open new research avenues, enabling incorporation of Huygens’ waveguide in the future integrated photonic platforms for both classical and quantum optical communication, computing and sensing applications. 

\section*{Funding Sources}

The project was funded by the (1) National Research Council Canada (NRC)'s (i) Postdoctoral Fellowship Program, (ii) Collaborative Science, Technology and Innovation Program (CSTIP) Small Teams (ST-R2-01-02), and by the (2) Deutsche Forschungsgemeinschaft (DFG, German Research Foundation, project number 448835038). M. Saad Bin-Alam acknowledges the support from the NRC's Postdoctoral Fellowship in Metamaterial Integrated Photonics.

\section*{Acknowledgements}

The authors thank Dr. Shurui Wang and Martin Vachon for their help in the experimental characterizations.

\section*{Supporting information}



The Supporting Information is available free of charge.

\begin{itemize}
  \item Multipole expansion and decomposition analysis; Forward directional cross-coupling; Resonance spectra, free spectral range (FSR), group index and group velocity dispersion; Resonance splitting and linewidth narrowing; Characteristics of the contra-directional coupler (CDC)-loaded racetrack resonators 
\end{itemize}

\section*{Associated Content}

A pre-print version of this manuscript is available at arXiv:2510.15845. \

\noindent Bin-Alam, M. S., Sirmaci, Y. D., Fernández-Hinestrosa, A., Zhang, J., Dolgaleva, K., Boyd, R. W., Luque-González, J. M., Pertsch, T., Staude, I., Schmid, J. H., Cheben, P., Directional and contra-directional coupling in Huygens' metawaveguide microring resonators. 2025, arXiv preprint arXiv:2510.15845. \url{https://doi.org/10.48550/arXiv.2510.15845} (March 27, 2026).

\bibliography{acs-template}

\newpage

\centering
\onecolumn
For Table of Contents Only
\begin{figure}[H]
\begin{centering}
\includegraphics[width=3.25in]{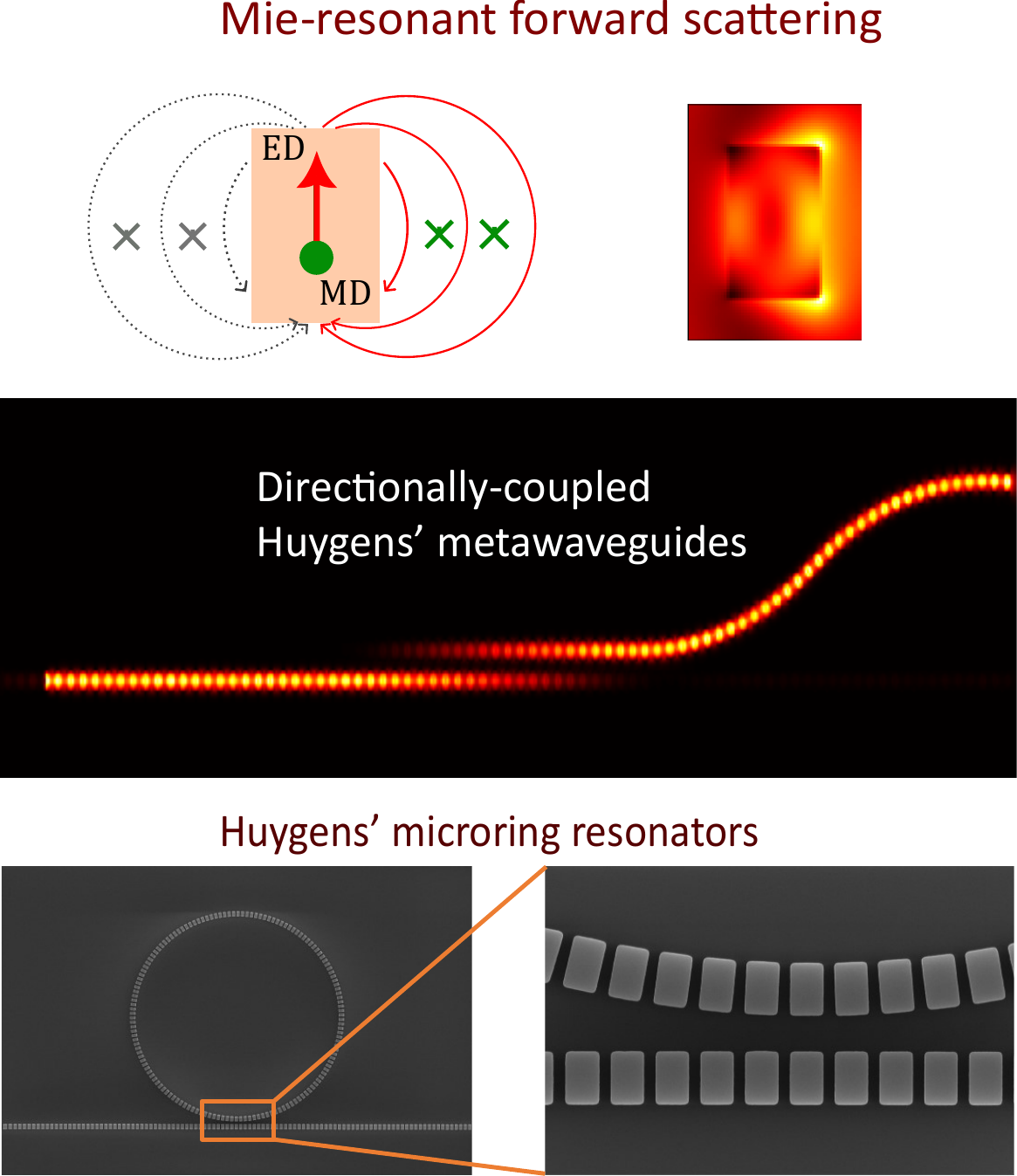}
\par\end{centering}
\end{figure}



  
  

\end{document}